\documentclass[aps,pre,twocolumn,showpacs,superscriptaddress,groupedaddress,nofootinbib]{revtex4-2} %preprint

\usepackage{graphicx}% Include figure files
\usepackage{dcolumn}% Align table columns on decimal point
\usepackage{bm}% bold math
\usepackage{color}
\usepackage{amsmath}
\usepackage[normalem]{ulem}
\newcommand{\tasd}{\overline{\delta^2(\tau,t)}}

\newcommand{\average}[1]{\left< #1 \right>}

\begin{document}

\title{Classification of anomalous diffusion in animal movement data using power spectral analysis}

\author{Ohad Vilk$^{a,b,c, *}$, Erez Aghion$^{d}$, Ran Nathan$^{b, c}$, Sivan Toledo$^{c, e}$, Ralf Metzler$^{f,*}$, Michael Assaf$^{a,f,}$}
\email{ Correspondence should be sent to:  ohad.vilk@mail.huji.ac.il, rmetzler@uni-potsdam.de, michael.assaf@mail.huji.ac.il}

 \affiliation{$^a$ Racah Institute of Physics, The Hebrew University of Jerusalem, Jerusalem, Israel,}
 \affiliation{$^b$ Movement Ecology Lab, Department of Ecology, Evolution and Behavior, Alexander Silberman Institute of Life Sciences, Faculty of Science, The Hebrew University of Jerusalem, Jerusalem, Israel,}
 \affiliation{$^c$ Minerva Center for Movement Ecology, The Hebrew University of Jerusalem, Jerusalem, Israel,}
 \affiliation{$^{d}$Departments of Physics and Chemistry, University of Massachusetts Boston, MA 02125, USA,}
 \affiliation{$^{e}$Blavatnik School of Computer Science, Tel-Aviv University, Israel.}
 \affiliation{$^{f}$Institute of Physics and Astronomy, University of Potsdam, Potsdam 14476, Germany}

\begin{abstract}
The field of movement ecology has seen a rapid increase in high-resolution data in recent years, leading to the development of numerous statistical and numerical methods to analyse relocation trajectories. Data are often collected at the level of the individual and for long periods that may encompass a range of behaviours. Here, we use the power spectral density (PSD) to characterise the random movement patterns of a black-winged kite (\textit{Elanus caeruleus}) and a white stork (\textit{Ciconia ciconia}). The tracks are first segmented and clustered into different behaviours (movement modes), and for each mode we measure the PSD and the ageing properties of the process. For the foraging kite we find $1/f$ noise, previously reported in ecological systems mainly in the context of population dynamics, but not for movement data. We further suggest plausible models for each of the behavioural modes by comparing both the measured PSD exponents and the distribution of the single-trajectory PSD to known theoretical results and simulations.
\end{abstract}

\maketitle
%23/03 01:59 am

\section{Introduction}
In natural stochastic phenomena, random fluctuations are often modelled via uncorrelated ``white noise", with finite mean and variance, %(often associated with Gaussian distributions)
%and an
%constant
%power spectrum density~\cite{gardiner1985handbook} pointing to
%equal noise
and with a fluctuation amplitude that is independent of the  frequency~\cite{gardiner1985handbook}.  However, a large class of natural systems exhibit so-called $1/f$ noise, considered to be an emergent property of correlations that extend across multiple temporal scales~\cite{mandelbrot1967some, voss1992evolution,csabai19941, ivanov20011, balandin2013low, moon2018intrinsic}. %\green{The term ``$1/f$" refers to  the amplitude of this noise, which grows as the reciprocal of the fluctuation frequency $f$.}
Traditionally, $1/f$ noise is defined in terms of the power spectral density (PSD). For an \textit{individual} \textit{trajectory},  the PSD of a $d$-dimensional path, $\textbf{x}(t) = (x_1(t), x_2(t), ..., x_{d}(t))$, as a function of time $t$, is defined in terms of the sum of squares of the Fourier transform over the $d$ individual components \cite{krapf2018power, krapf2019spectral}
\begin{equation} \label{PSD_def_d}
    S_d(f, t_m) = \frac{1}{t_m} \sum_{j = 1}^{d} \left|\int_0^{t_m} e^{i f t} x_j(t) dt\right|^2.
\end{equation}
Here, $t_m$ is the measurement time of the process and $f$ the frequency.
``Coloured noise'' corresponds to the asymptotic scaling $\average{S(f)} \propto 1/f^\beta$ of the averaged PSD at low frequencies (see below), where angular brackets imply ensemble averaging, and  $0\leq\beta\leq2$  \cite{mandelbrot2002gaussian}.  %\red{[[Erez: Think on $0\leq\beta<1$, and $\beta>2$.]]}
Notably, we here focus on the regime $1\leq\beta\leq2$ where $S(f)$ is non-integrable as $\int_0^\infty S(f)df \sim \int_0^\infty 1/f^\beta df \rightarrow\infty$. Naturally, low frequencies that cause such divergence are associated  with infinitely long correlation and measurement times. %\blue{[[Erez, I think this issue of long measurement times that cause the catastrophe is still not clear enough. For instance: in any real system $t_m$ is finite and so $f$ is bounded. Should we add a sentence about this?]]} \red{[[-- That's right, this is the whole story behind the idea of the aging power spectrum, explained in the next paragraph (originally this was a single paragraph). It is precisely why Eq. (2) was introduced, in order to resolve the ``paradox" (as it's in-fact written right below... I am adding some more explanations in blue in this paragraph, check it out?)]]}
This non-integrability is often referred to as ``the infrared catastrophe", as it implies that infinite energy is stored in the low frequency fluctuations, even in a bounded process, which is evidently not physical~\cite{mandelbrot1967some,niemann2013fluctuations, sadegh20141}. Nevertheless, experimentally, power-law shaped PSDs attributed to $1/f$ noise have been inferred from finite-time observations in systems ranging from semiconductor devices and metal films \cite{mandelbrot1967some, dutta1981low} to earthquakes \cite{sornette1989self} and human cognition \cite{gilden19951}.
 %\blue{Generally, in such this phenomenon may lead to anomalous transport, with  non-Gaussian distribution~\cite{}. [[I am not sure about placing this sentence here, what do you think?]]}

In recent years, the ``paradox" of the divergent power spectrum was  resolved by showing that at finite measurement times the exponent $\beta$ is not sufficient to characterise all properties of the PSD. Specifically, to physically interpret the apparent divergence at small $f$, one should use the non-stationary, \textit{ageing power spectrum}, of the form~\cite{leibovich2015aging,leibovich2016aging}
\begin{equation} \label{S_scaling_form}
    \average{S(f,t_m)} \propto t_m^{-z}f^{-\beta},
\end{equation}
where $z$ is called the \textit{ageing exponent}. This form entails that the divergent nature of the spectrum expected at infinite times manifests itself through the (finite-time) finite value of a non-stationary power spectrum~\cite{leibovich2015aging,leibovich2016aging,leibovich20171}. This ageing pattern has been measured in a range of processes~\cite{bak59self,jensen1991fractal,banerjee2006self,matthaeus1986low,jin20211}, including intermittent quantum dots \cite{sadegh20141}, telomeres in the nucleus of cells \cite{krapf2019spectral} and the motion of membrane
proteins in living cells \cite{FoxPowerSpectrum}.  %\red{[[Erez, could you expand the discussion on $1/f^\beta$ noise following our last conversation? Also, if you are aware of examples I missed here feel free to add them.]]}

In ecology, $1/f$ noise is primarily observed in population dynamics and thought to account for the variability and autocorrelation of such ecological time-series \cite{halley2004increasing}. For example, the variability of the population density in multiple species has been shown to increase over time, a phenomena that can be accounted for by $1/f$ noise \cite{pimm1988variability}. Rapid increase in high-resolution movement data has led to the development of numerous statistical and numerical methods~\cite{gurarie2016animal, seidel2018ecological}, as well as analytical methods~\cite{vilk2021phase} to analyse relocation trajectories. Data is often collected for multiple individuals and for long periods that encompass a range of behaviours, from local searches within a bounded patch, to long migratory flights~\cite{ran2022BigData}. Here, the PSD is also used in the analysis of movement data, mainly to detect frequencies that account for the variability across a movement path (see, e.g., \cite{riotte2013periodicity}). However, to the best of our knowledge, neither the PSD nor its ageing effects have been used to model and classify the noise of movement paths in movement ecology.
Specifically, ageing can have important implications on the process, entailing if it is stationary and ergodic, which in turn affects the ability to average over common observables (e.g., net and squared displacements, as well as velocity)~\cite{vilk2021ergodicity, mangalam2021point}.

In processes with $1/f^2$ noise (corresponding to Brownian noise), finite-time effects in the power spectrum were shown to decay in the limit $t\rightarrow\infty$, and averaging over an ensemble of independent trajectories, the value measured numerically from finite paths converges to a single function $\langle S(f)\rangle$ as the path length is increased~\cite{krapf2018power}. However, having an ageing power spectrum with $1\leq\beta<2$, the dependence of the PSD on both $f$ and $t_m$ always persists. To account for the time dependence, periodograms are often used~\cite{peron2016uncovering}.
Additionally, several recent studies have focused on the PSD of individual trajectories, which enable data analysis of experimental systems even where only few trajectories are available. It is commonly the case that the PSD remains very stable across trajectories, as was explicitly shown for Brownian motion (BM) and fractional Brownian motion (FBM) \cite{krapf2018power, krapf2019spectral}. %\red{[[Erez, are there other reasons to consider individual trajectories and not only the average, aside from what I wrote here?]]}

Notably, the study of the ageing properties of PSD is complementary to the study of the ensemble-averaged (EA) and ensemble-averaged time-averaged (EA-TA) mean squared displacement (MSD) \cite{metzler2014anomalous}. The EA-MSD, $\average{\mathbf{x}^2(t)}$, is defined as the squared displacement of an individual's position with respect to a reference position, averaged over an \textit{ensemble} of movement paths. The EA-TA-MSD is given by averaging over the squared displacement performed during a time lag $\tau$, and then averaging again over the ensemble \cite{he2008random, barkai2012single}
\begin{equation} \label{TASD}
\average{\tasd} = \average{\frac{1}{t - \tau}\int_0^{t - \tau} [\mathbf{x}(t' + \tau) - \mathbf{x}(t')]^2 dt' }.
\end{equation}
The EA- and EA-TA-MSD are commonly used to study correlations and ergodicity in anomalous systems~\cite{metzler2014anomalous}, and when discrepancy appears the process is said to exhibit weak ergodicity breaking, see e.g.,~\cite{bouchaud,burov2011single}. An important observable characteristic for different stochastic processes, is the scatter of amplitudes of the TA-MSD at a given time lag \cite{he2008random, metzler2014anomalous}
%\blue{[[Again, there is a point here which we should talk about, to make it more clear in the text: Are you referring to the difference between EA and TA power-spectrum, like in~\cite{leibovich2015aging,leibovich2016aging}? This isn't completely clear now, but we can sort it out easily. ]]}

In this study, we perform an empirical analysis of the PSD of movement tracks of a kite and a stork. These animals are chosen as their recorded tracks are of different spatiotemporal scales: high resolution data for the kite (0.25 Hz) and long track duration for the migrating stork ($>$ 8 years). Additionally, for both animals the tracks are known to encompass a range of different behaviours (see below). As animal movement typically varies across spatiotemporal scales, and strongly depends on seasonality as well as behavioural mode \cite{nathan2008movement}, we focus here on interpreting different movement modes of an animal, shown to each have different (non-)stationary properties. %We discuss our results in relation to known theoretical results for a number of models. Here,
We further analyse the single-trajectory PSD, which can vary depending on the physical process and on the dimension $d$. Finally, we find significant ageing effects when the initiation time of the process differs from the initial measurement time. %, is shown to alter properties of the PSD and the TA-MSD.
To interpret our findings we use simulations and theoretical results developed in recent works for several well-known physical models, BM \cite{norton2003fundamentals, krapf2018power}, FBM \cite{krapf2019spectral} and the subordination of FBM to continuous-time random walks (CTRW)~\cite{leibovich2016aging, FoxPowerSpectrum} (details below). %\blue{The latter, is a process built from a series of jumps (in continuous time), separated by random sojourn times with a diverging mean value, and the final position of the particle is determined by the sum of the random number jump displacements in the time interval $[0,t]$~\cite{klafter2011first} (see explanations, below).}
These models allow us to infer properties of the observed movement paths, particularly to distinguish sub- from super-diffusion, namely diffusion where then MSD grows slower or faster than linear in time, and to quantify its non-stationary properties.

\section{Materials and methods}
\subsection{Data collection}

\textbf{Black winged kite}.
An individual black-winged kite (\textit{Elanus caeruleus}), residing in the Hula Valley, Israel, was tracked using ATLAS, an innovative reverse-GPS system. ATLAS localises extremely light-weight, low-cost tags~\cite{toledo2020cognitive}, where each tag transmits a distinct radio signal which is detected by a network of base-stations distributed in the study area. Tag localisation is computed using nanosecond-scale differences in signal time-of-arrival to each station, alleviating the need to retrieve tags or have power-consuming remote-download capabilities~\cite{toledo2020cognitive, vilk2021ergodicity}. The kite was tracked for 164 consecutive days in years 2019-2020, with a mostly constant tracking frequency of 0.25 Hz, see Ref. \cite{vilk2021ergodicity} for more details. Our analysis is limited to data collected during the activity hours (omitting the nights for the diurnal kite) and we further excluded data collected in proximity to the observed nest to focus on local search behaviour.

\textbf{White stork}.
An adult white stork (\textit{Ciconia ciconia}) was tracked between May 2012 and July 2020 using the configuration described in  Ref.~\cite{rotics2016challenges}. Here the GPS location and speed were recorded during the day at a frequency of 1/300 Hz when solar recharge was high (92\% of the time), and at 1/1200 Hz otherwise.
We omit days with lower frequency ($<1\%$ of tracked days) and only include localisations that occur between the first recorded velocity of $>4$ m/s~\cite{vilk2021unravelling}.

% \textbf{Eurasian griffon vulture}.
% An Eurasian griffon vulture (\textit{Gyps fulvus}, Hablizl 1783) was tracked in Israel and surrounding countries with high-resolution global positioning system (GPS), between October 2012 and October 2015. The 90-g GPS transmitters (E-Obs GmbH; Munich, Germany) were fitted in a backpack configuration and set to a 13 h duty cycle, between 7:00 a.m. and 8:00 p.m. to correspond with the vulture's activity pattern~\cite{harel2016decision}. Localisations were optimally recorded at a frequency of 1/600 Hz (73\% of the time for this vulture), or 1/1200 Hz (23 \% of the time). Vulture days tracked at lower frequencies were omitted from this study. %Transmitting efforts were approved by the Israel Nature and Parks Authority and were in accordance with the ethics guidelines of the Hebrew University of Israel (NS-07-11063-2).
% See Ref.~\cite{harel2016decision} for more details.

% As the time of the vulture's departure from the nest can drastically vary between different days, for each tracking day we omit localisations that occur prior to the vulture's first detected movement~\cite{harel2016decision}. This was done by removing all localisations that occur before the first recorded velocity of $>4$ m/s during the same day.

\subsection{Theoretical models}
In this work we focus on empirical evidence for $1/f$ noise and ageing of the PSD in ecological movement data. In several cases it is possible to relate our findings to known physical models. Here, we list those models, for which analytical results for the PSD have been reported in recent years.
%\red{[[Erez, please read this part carefully. Feel free to add anything that is missing, including references.]]}

\subsubsection{Brownian motion}
For an overdamped, one-dimensional Brownian trajectory $x(t)$, with $t \in [0, t_m]$, the dynamics is given by the Langevin equation  \cite{coffey2012langevin}
\begin{equation} \label{BM_lang}
    \frac{dx(t)}{dt} = \xi(t)
\end{equation}
where $\xi(t)$ is a Gaussian, delta-correlated white noise term with zero mean and $\average{\xi(t)\xi(t')} = 2 D \delta(t-t')$,  and $D$ is the diffusion constant. For a particle governed by Eq. \eqref{BM_lang}, the EA- and EA-TA-MSD, Eq. \eqref{TASD}, are respectively given by 
$\average{\tasd} = 2 D \tau$
and
$\average{x^2(t)} = 2 D t$~\cite{balandin2013low, barkai2012single}. As both show identical scaling with time, the process is ergodic in the (weaker) Boltzmann-Khinchin sense of equality between (long) time and ensemble averages \cite{metzler2014anomalous}.
For this process, the mean of the PSD follows \cite{norton2003fundamentals}
\begin{equation} \label{PSD_BM}
    \average{S(f)} \simeq \frac{4D}{f^2},
\end{equation}
entailing $\beta = 2$  and that the amplitude of the PSD is linear in $D$ without explicit dependence on $t_m$. Here, the non-integrability of the noise is generally solved by considering an unbounded process, however it does demonstrate the significant difference between ``white" noise and processes where the noise is Brownian (sometimes called ``red" or ``brown" noise \cite{halley2004increasing, krapf2018power}). %\red{[[Note that in the introduction we offer a different explanation for why this isn't paradoxical. I guess that one was suggested by Ralf? -- for consistency we can delete the explanation I gave here in this case, and just say that this is called "Brownian noise" or "red noise", and add the citations.]]}.
% \red{[[I prefer following phrasing, below, but it's not "mandatory": "\blue{Such noise, which is generated by the  integral of white noise, is sometimes called ``red" or ``brown" noise \cite{halley2004increasing, krapf2018power}. Its  non-integrability, however,  is generally not thought of as  paradoxical, since it is mostly considered in unbounded processes.}"]]}

For any finite $t_m$ the single-trajectory PSD will fluctuate from one realisation to the next. The distribution of the single-trajectory PSD [Eq.~\eqref{PSD_def_d}] around its ensemble average is then quantified by the amplitude \cite{krapf2018power}
\begin{equation} \label{Ad_def}
    \frac{S_d(f, t_m)}{\average{S_d(f, t_m)}} \equiv A_d(f, t_m)
\end{equation}
where $A_d$ is a random number that depends on the choice of the stochastic model (see examples below), and can also depend on the dimension $d$, the frequency $f$, and the measurement time $t_m$. The distribution of $A_d$ is highly informative since it is model specific as shown analytically for a number of processes \cite{krapf2018power, sposini2019single, krapf2019spectral, metzler2019brownian}. In particular, it is important in studies where only few paths are available, as it relates the single PSD to its average.
For BM, the distribution of the single-trajectory PSD around the mean \eqref{PSD_BM} is given by Eq. \eqref{Ad_def}, with \cite{krapf2018power}
\begin{equation} \label{PA_BM}
    P(A_d = A) = \frac{2 \sqrt{\pi}A^{(d-1)/2}}{\sqrt{3} \Gamma(d/2)} e^{-\frac{4}{3}A}I_{(d-1)/2}\left(\frac{2}{3}A\right),
\end{equation}
where $I_{\xi}(\cdot)$ is the modified Bessel function of the first kind and $\Gamma(\cdot)$ is the Gamma function. Notably, the shape of the distribution depends on the dimensionality, where dimensions lower than the embedding space of the process can be achieved by projection or component-wise measurement \cite{krapf2018power,krapf2019spectral}.

\subsubsection{Fractional Brownian motion}
\label{SubSecFBM}
As in the case of BM, FBM can also be defined in terms of the Langevin equation \eqref{BM_lang},
%\red{[[I think this is wrong... needs to have a memory kernel, or fractional derivative, no?]]}
replacing the delta-correlated noise term $\xi(t)$ with zero-mean, power law correlated fractional Gaussian noise $\xi_{\text{fGn}}(t)$ defined by \cite{metzler2014anomalous}
\begin{equation} \label{xi_fgn}
    \average{\xi_{\text{fGn}}(t)\xi_{\text{fGn}}(t')} \sim 2H(2H-1)|t - t'|^{2(H - 1)}
\end{equation}
% \blue{at large $t\gg t'$}.
for $t \neq t'$.
Here, $H$ is termed the Hurst exponent and, similarly to the case of BM, the process is ergodic with $\average{\tasd} = 2 \tilde{D} \tau^{2H}$
and
$\average{x^2(t)} = 2 \tilde{D} t^{2H}$, where $\tilde{D}$ is an effective diffusion constant~\cite{deng2009ergodic, metzler2014anomalous,wang2020fractional}. In accordance with the definition of $\xi_{\text{fGn}}(t)$, for $H > 0.5$ the noise is persistent leading to a positively correlated process while for $H< 0.5$ it is antipersistent leading to a negatively correlated process.
For FBM and other stationary (in increments) processes, the PSD is written in terms of the EA autocorrelation function $C_{EA}(\tau) = \average{x(t) x (t + \tau)}$ by the Wiener-Khinchin theorem
\begin{equation}
    \average{S(f, \infty)} = \int_{-\infty}^{\infty} e^{i f \tau}C_{EA}(\tau) d\tau.
\end{equation}
Here, $C_{EA}$ and the time averaged (TA) autocorrelation function, defined by
\begin{equation}
    C_{TA}(t_m, \tau) = \frac{1}{t_m - \tau} \int_0^{t_m - \tau} x(t) x(t+ \tau) dt,
\end{equation}
are identical at long times. %\red{[[Erez, I feel like this part is lacking, should we provide more details here? See also the subordinate FBM model below.]]} \blue{[[Which type of additional detail? Why is the Wiener Khinchin defined only here actually, only in the context of FBM? It is general...]]}
In the limit of long measurement times it was shown that for FBM \cite{krapf2019spectral}
\begin{equation} \label{PSD_FBM}
    \average{S(k, t_m)} \sim\begin{cases}
     t_m^{2 H - 1}f^{-2} & H > 1/2, \\
     f^{-1-2H} & H < 1/2,
    \end{cases}
\end{equation}
i.e., for persistent FBM the PSD is ageing, while it is not for subdiffusive FBM.
%In the case of negative correlations ($\mathcal{H} < 1/2$) the process displays $1/f$ noise but no ageing, and for positive correlations if displays ageing but no $1/f$ noise.
Here, it is also possible to obtain the distribution of the single-trajectory PSD around the mean; for example, for superdiffusive FBM ($H > 1/2$) in $d$-dimensions it was found that \cite{krapf2019spectral}
\begin{equation} \label{PA_FBM_Super}
    P(A_d = A) = \frac{A^{d/2 - 1}}{2^{d/2}\Gamma(d/2)}e^{-A/2},
\end{equation}
see Ref. \cite{krapf2019spectral} for more details and results. Here, the distributions $P(A_d)$ are in general different from those of BM, compare Eq. \eqref{PA_BM}.

\subsubsection{Subordinated FBM}
\label{SubSecSubordinatedFBM}
% We employ the subordinated FBM framework suggested by \cite{FoxPowerSpectrum}.
Here we consider a process in discrete time steps, expressed by the number of jumps $n = 1, 2, 3, \dots$, such that the autocorrelation function is given by
\begin{equation} \label{FBMdef}
    \average{x_n x_{n + \Delta n}} = \Delta x^2 [n^{2\mathcal{H}} + (n + \Delta n)^{2\mathcal{H}} - n^{2\mathcal{H}}]
\end{equation}
where $\Delta x$ is a scaling parameter. Furthermore, $\mathcal{H}$ is analogous to the Hurst exponent of the FBM, although here it is not directly related to the scaling of the EA-MSD on time (see below). In this process, the discrete-time FBM \eqref{FBMdef} is subordinated to the operational time of the CTRW,  %\footnote{If the operation time $t$ is linear with the discrete time $n$, Eq. \ref{FBMdef} can be shown to be equivalent to the Langevin description of the FBM, see Eq. \eqref{xi_fgn} above, and the dynamics are not subordinated to CTRW.},
such that after each FBM step, the particle is immobilised for a random waiting time $\tau$ drawn from a fat-tailed distribution \cite{klafter2011first}
\begin{equation} \label{psi_tau}
    \psi(\tau) \simeq \tau^{-(1+\alpha)},
\end{equation}
with $0<\alpha<1$, such that $\langle\tau\rangle$ diverges. By combining the correlation structure of the jump sizes of FBM, with the unbounded waiting times of CTRW, this process effectively distinguishes the "operative clock", defined by the sequence of jump events themselves, from the "real" clock defined by $t$, thus allowing for both temporal correlations as well as ageing effects (as in individual processes, FBM only allows for the former, while CTRW allows for the latter~\footnote{When $\alpha>1$, since the mean sojourn time is finite, the operation time $t$ becomes linear with the discrete time $n$, and the process is equivalent to standard FBM, see Sec. \ref{SubSecFBM}.}.) Waiting time distributions similar to Eq.~\eqref{psi_tau} have been reported in multiple empirical systems, e.g., \cite{brokmann2003statistical, weigel2011ergodic, song2018neuronal, fernandez2020diffusion, vilk2021ergodicity}.

As subordinated FBM is a non-stationary process, $C_{EA}$ and $C_{TA}$ are not identical, and the Wiener-Khinchin theorem no longer applies. Similar observations in other processes have led to the development of the ageing Wiener-Khinchin theorem in recent years \cite{leibovich2015aging, leibovich2016aging}, which enabled analytical calculations of the PSD of both stationary and non-stationary processes, relating the exponents $\beta$ and $z$ to underlying physical processes, see, e.g., \cite{leibovich2016aging, krapf2019spectral, metzler2019brownian, FoxPowerSpectrum, sposini2019single, sposini2020universal}.
In particular, for subordinated FBM the PSD has been rigorously computed in Ref. \cite{FoxPowerSpectrum}. The leading term of the spectrum was found to again differ between positively and negatively correlated increments as \cite{FoxPowerSpectrum}
\begin{equation} \label{PSD_FBM_CTRW}
    \average{S(k, t_m)} \sim \begin{cases}
     t_m^{2\alpha \mathcal{H} - 1}f^{-2} & \mathcal{H} > 1/2, \\
    t_m^{\alpha - 1}f^{-2+\alpha(1-2\mathcal{H})} & \mathcal{H} < 1/2.
    \end{cases}
\end{equation}
Note that for $\alpha = 1$, Eq. \eqref{PSD_FBM} is retrieved, i.e., the process corresponds to FBM. In Eq. \eqref{PSD_FBM_CTRW}, for positively correlated increments $\mathcal{H} >1/2$, the process does not exhibit $1/f$ noise (i.e., $\beta = 2$, as is the case for BM, see the discussion in \cite{krapf2019spectral}); however, the ageing properties depend on $\alpha$: for $\alpha \mathcal{H} > 1/2$ the spectrum increases with the measurement time $t_m$ as in superdiffusive FBM \cite{krapf2019spectral}, while for $\alpha \mathcal{H} < 1/2$ the spectrum amplitude  decreases with $t_m$.
In contrast, for anticorrelated increments in terms of the number of jumps $n$ ($\mathcal{H} < 1/2$), for any $\alpha < 1$ the spectrum displays $1/f$ noise ($\beta < 2$) and ageing which decreases with the measurement time ($z < 0$).
Below we compare these theoretical predictions to the recorded movement paths of a kite and a stork, for which we can obtain the values of $\alpha $ and $\mathcal{H}$ and infer the non-stationary properties of different movement modes.

Importantly, the values of $\alpha$ and $\mathcal{H}$ in this model can also be obtained from independent analysis of the EA-TA-MSD, Eq. \eqref{TASD}. For the subordinated process, when $0 < \alpha < 1$ the EA-TA-MSD is given by  \cite{meroz2010subdiffusion}
\begin{equation} \label{MSD_FBM_CTRW}
%   \average{x^2(t)} \sim t^{2 \alpha H},  \quad \quad
   \average{\tasd} \sim \frac{\tau^{1 - \alpha + 2\alpha \mathcal{H} }}{t_m^{1-\alpha}}.
\end{equation}
The EA-TA-MSD was found for our empirical data (kite and stork) in a recent study \cite{vilk2021unravelling}; however, the MSD (and in particular the EA-MSD) is typically a noisy observable \cite{FoxPowerSpectrum}, and in some cases it was not possible to perform adequate power-law fits. Thus, the analysis of the PSD can be highly valuable when comparing the data to theoretical processes, as shown below.
% Notably, the long-time scaling of the EA-TA-MSD has also been written in previous studies in terms of the Joseph, Noah and Moses exponents in the following way \cite{meyer2018anomalous,aghion2021moses,zamani2021anomalous}
% \begin{equation}  \label{TASD_exponents}
%     \average{\tasd} \propto \frac{\tau^{2J}}{t^{2 - 2L-2M}}.
% \end{equation}
% By comparing Eqs. \eqref{MSD_FBM_CTRW} and \eqref{TASD_exponents} it is possible to obtain the parameters $\mathcal{H}$ and $\alpha$ of the subordinate FBM directly from the scaling exponents found in a previous study \cite{vilk2021unravelling}.
% \subsubsection{Scaled Brownian motion}
% \red{Can be added based on \cite{sposini2019single}, but not sure it is necessary.}
Finally, as the theoretical distribution of the single-trajectory PSD is yet unknown for subordinated FBM, we compare our results to stochastic simulations of the process~\footnote{Here, the increments of the FBM, in discrete time $n$, are obtained using the python function fgn from the package fbm (Fractional Gaussian noise -- fGn -- is the increment process of the FBM). The increments, $\delta r_n$, are projected to two dimensions by generating a random number $\theta \in [ -\pi,\pi]$ and setting $\delta x_n =  \delta r_n \cos\theta$ and $\delta y_n =  \delta r_n\sin\theta$. The FBM in two dimensions is given by the cumulative sums $x_n = \sum_{m = 0}^{n}\delta x_m$ and $y_n = \sum_{m = 0}^{n}\delta y_m$. The times between (the discrete-time) steps are then drawn from a Pareto distribution, Eq. \eqref{psi_tau}. Notably, there are also other ways to simulate FBM for $d = 2$. We are currently studying other simulation methods and how they will affect our results.}.

\begin{figure*}[t]
\centering
\includegraphics[width=0.7\linewidth]{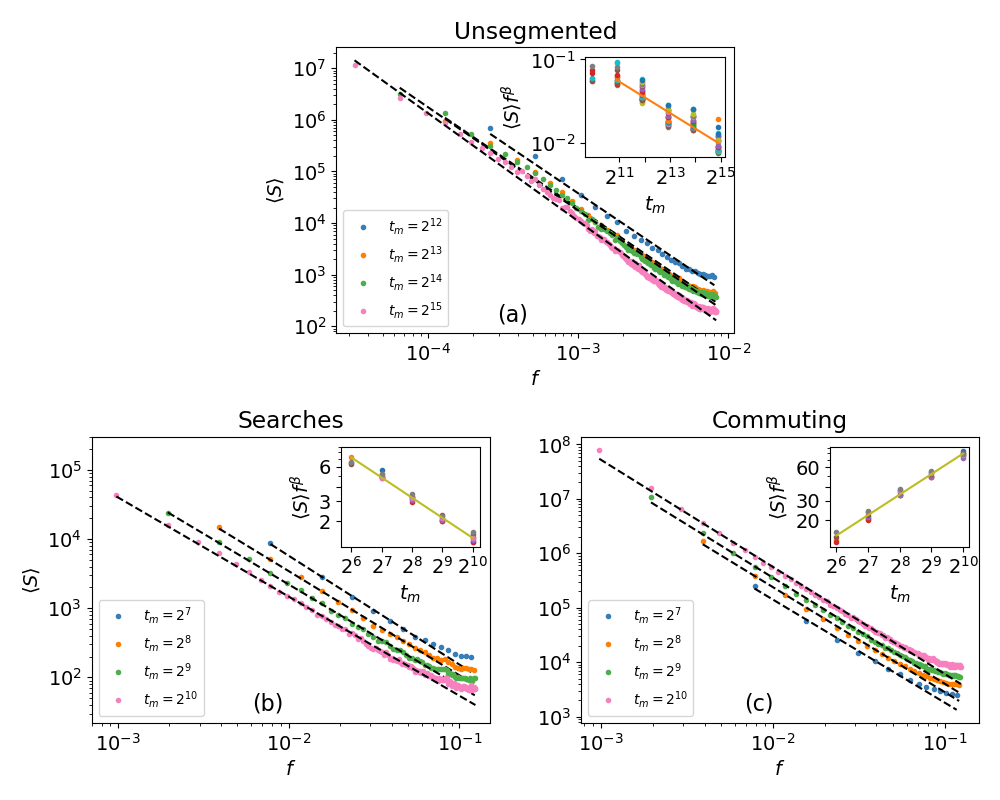}
    \caption{Averaged PSD, $\average{S(f, t_m)}$, of empirical trajectories of a kite for (a) unsegmented daily movement trajectories, compared to independent analysis of (b) area restricted searches and (c) commutes, for different measurement times (coloured dots, see legends). The averaged PSD is fitted to a power law (black dashed lines). In the insets we plot the amplitudes (scaled by $f^\beta$) as a function of $t_m$ for different frequencies (coloured dots), with a fitted scaling (solid line). For searches (b) we find $\average{S} \sim f^{-1.50}t_m^{-0.57}$, for commutes (c) $\average{S} \sim f^{-2}t_m^{0.60}$, while for unsegmented tracks (a) $\average{S} \sim f^{-2.10}t_m^{-0.66}$.
    }
 \label{fig:kite_analysis}
\end{figure*}

\section{Results}
\subsection{Kite} \label{sec:resultskite}
Similarly to Ref.~\cite{vilk2021ergodicity}, the kite's tracks are segmented into two behavioural modes: local searches (area restricted search) and commutes (directed flights between local searches). Localisations were segmented by detecting switching points in the data -- distinct points in which the bird switches between the two behaviours~\cite{benhamou2014scales}. Switching points were detected using spatiotemporal criteria segmentation, such that localisations that are in proximity to one another both in space and time, were segmented together \cite{gurarie2016animal}.
In accordance with the conclusions of Ref.~\cite{vilk2021ergodicity} we independently analysed the time series ensemble that represents instances of searches and the time series ensemble that represents commutes. Below, we compare these analyses to an unsegmented ensemble of daily tracks, showing that ecological knowledge of the underlying processes is crucial to correctly identify the noise properties of the data.

% In recent works \cite{vilk2021ergodicity, vilk2021unravelling} these ensembles have been analysed and the focus has been on the EA- and EA-TA-MSD. During area restricted search (Fig. \ref{fig:kite_analysis}a-b), the process is subdiffusive, nonergodic and can be modelled as a confined CTRW, as shown in Ref. \cite{vilk2021ergodicity}.
% For searches, an analysis of the MSD (Fig.  \ref{fig:kite_analysis}a) shows that $\average{x^2(t)} \sim t^{0.1}$ and $\average{\tasd} \sim \tau^{0.56}$. This discrepancy between time-averaging and ensemble-averaging indicated ergodicity breaking \cite{metzler2014anomalous}. To account for this effect, it has been suggested that   suggesting that  $H \simeq 0.09$ and $\alpha \simeq 0.53$ (compare Eq. \eqref{MSD_FBM_CTRW}).
For searches, the PSD  (Fig.~\ref{fig:kite_analysis}b) displays $1/f^\beta$ noise with $\average{S} \sim f^{-1.50}t_m^{-0.57}$. These results are consistent with subdiffusive subordinated FBM with $\alpha = 0.43 \pm 0.04$ and $\mathcal{H} \simeq 0$, see Eq. \eqref{PSD_FBM_CTRW}. Notably, the value of $\mathcal{H}$ is close to zero and the error is relatively large, thus an exact estimate of its value is hard to achieve.
An analysis of the MSD for the same ensemble of searches produces $\average{\tasd} \sim  \tau^{0.55} t^{-0.55}$, which is consistent with $\alpha = 0.45 \pm 0.03$ and $\mathcal{H} \simeq 0$, see Eq. \eqref{MSD_FBM_CTRW}.
Here, both the analysis of the MSD and of the PSD suggest that subordinated FBM is a plausible model for the movement within search grounds of this kite.
In contrast, for the commutes (Fig. \ref{fig:kite_analysis}c) we find that $\average{S} \sim f^{-2} t_m^{0.60}$, consistent with (ergodic) superdiffusive FBM 
%\red{\sout{(i.e., $\alpha >1$)}} 
with $H = 0.80 \pm 0.03$, see Eq. \eqref{PSD_FBM}. Here, analysis of the EA-TA-MSD gives a scaling of $\average{\tasd} \sim \tau^{1.66}$, suggesting $H = 0.83 \pm 0.03$, in agreement with the analysis of the PSD.

We compare these results to the PSD of unsegmented daily trajectories (Fig. \ref{fig:kite_analysis}a). Here, the trajectories include both the commutes and searches, each occurring at different temporal and spatial scales. We find that $\beta = 2.10$, however there is strong ageing in the process as $\average{S} \sim t_m^{-0.66}$. Notably, the $1/f$ noise found during searches is completely skewed due to the lack of segmentation, and the process is most likely a mixture of (at least) two different processes, making these results hard to interpret. Additionally, the dependence of the PSD on $t_m$ (Fig. \ref{fig:kite_analysis}a, inset) does not provide a clear scaling form when normalised by $f^{-\beta}$ for different $f$ as is the case for searches and commutes (Fig. \ref{fig:kite_analysis}b-c, insets). These findings highlight the importance of correctly accounting for different movement phases, allowing us to better interpret the $1/f$ noise in the data. %A plausible model here may be CTRW with $\alpha = 0.34$ and $H \simeq 0.5$, or SBM \red{details to be completed, I think there are no long waiting times at this scale so I would go for SBM) }.

Some discussion on stationarity is in order here. For searches, the longer we observe the system, the smaller the amplitude $A$ of the PSD becomes (Fig. \ref{fig:kite_analysis}b, inset). This occurs because the longer we track the searching bird, the more likely we are to find it "trapped" at a specific location for long periods; thus, the rate at which the animal moves is significantly reduced and the noise levels decrease accordingly \cite{sadegh20141}. In a recent work we have shown that this effect originates from long waiting times during searches \cite{vilk2021ergodicity}. The scaling exponent which is proportional to $\alpha$ [see Eq. \eqref{PSD_FBM_CTRW}] allows for yet another way to quantify this effect.
In contrast, for commutes the PSD \textit{increases} with the measurement time (Fig. \ref{fig:kite_analysis}c, inset). This process can be modelled with the use of FBM (i.e., $\alpha > 1$) and is Gaussian and ergodic in nature. As these flights are highly correlated, persistent movement at increasingly long measurement times increases the amplitude of the PSD -- hence the growth of the PSD.

\begin{figure}[t]
\centering
\includegraphics[width=1.\linewidth]{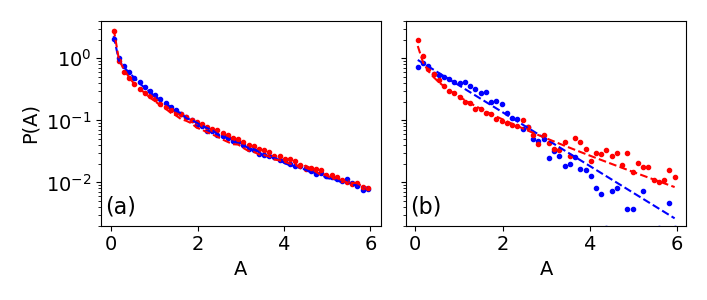}
    \caption{Distribution of single trajectory PSD of empirical trajectories of a kite in $d = 1$ (red points) and $d = 2$ (blue points) dimensions, for (a) searches and (b) commutes. In (a) the distributions are compared to simulations of subordinated FBM with $\alpha = 0.45$ and $\mathcal{H} = 0.01$, while in (b) they are compared to the theory for superdiffusive FBM, Eq. \eqref{PA_FBM_Super} for $d = 1, 2$ (dashed red and blue lines respectively). Notably, for the case of  subordinated FBM the theoretical distributions are yet unknown.}
 \label{fig:kite_analysis_PA}
\end{figure}

\begin{figure*}[t]
\centering
\includegraphics[width=0.85\linewidth]{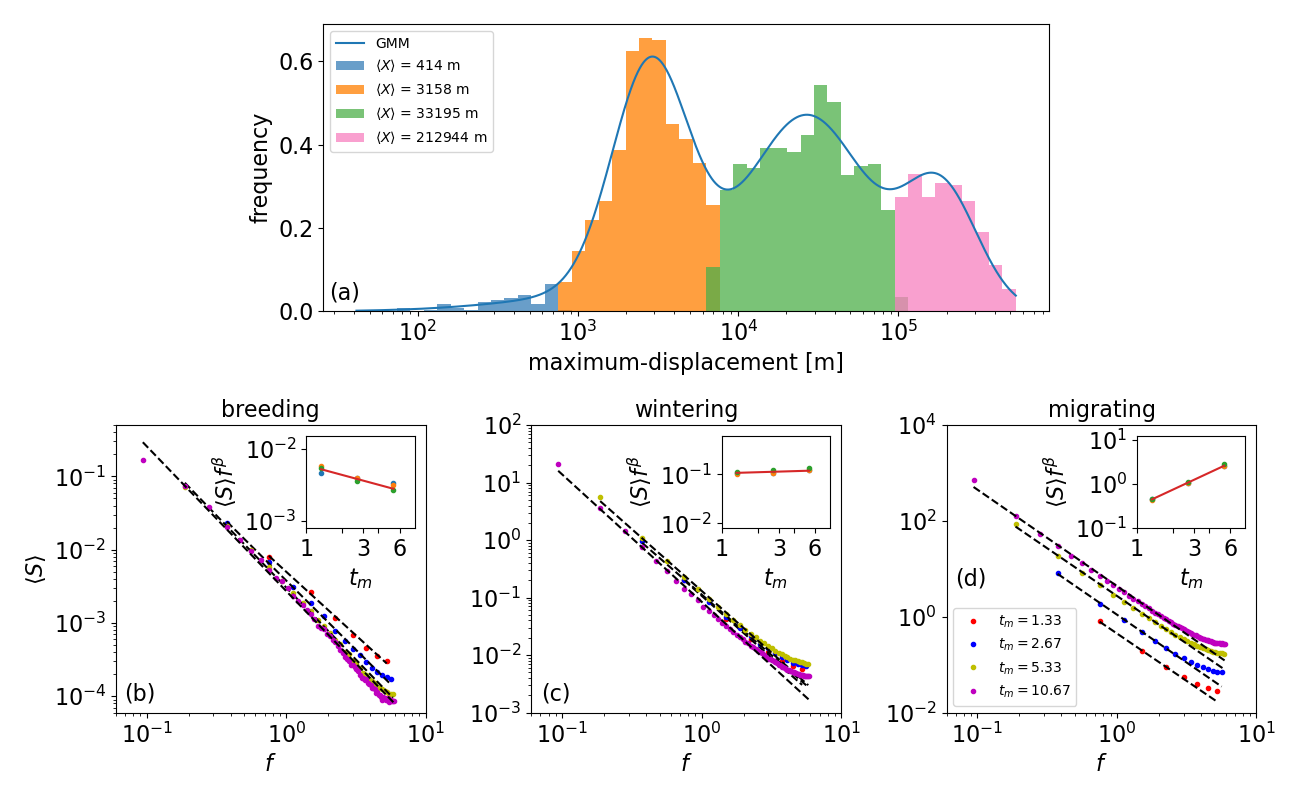}
 \caption{(a) Behavioural modes of a white stork. The clustering is performed using a Gaussian mixture model \cite{scikitlearn}, identifying four clusters, three of which account for more than 18\% each of the paths and thus represent a large subset of daily paths. The cluster identified on the far left includes $<3$\% of the data and is thus discarded from any analysis. (b-d) Average PSD for different measurement times [coloured dots, see legend in (d)] for each of the modes: (b) breeding, (c) wintering and (d) migrating. In the insets we plot the amplitudes (scaled by $f^\beta$) as a function of $t_m$ for different frequencies (coloured dots), with a fit (solid line).}
 \label{fig:storksGMM}
\end{figure*}

Finally, in Fig. \ref{fig:kite_analysis_PA} we plot the distribution of $A_d$ given by Eq. \eqref{Ad_def} for searches and commutes in $d=$1 and $d=$2 dimensions. For searches (Fig. \ref{fig:kite_analysis_PA}a), we compare these distributions to simulation results for subordinated FBM with $\alpha = 0.45$ and $\mathcal{H} = 0.01$, which are consistent with the results obtained above. For both dimensions we find excellent agreement between the simulated and observed distributions. Interestingly, we find that for searches, the empirical distribution does not strongly differ between $d = 1$ and $d = 2$. Here, for subordinated FBM a theoretical distribution for $A$ is yet unknown. In comparison, for the commutes (Fig. \ref{fig:kite_analysis_PA}b) the distributions are in good agreement with the theory given by Eq. \eqref{PA_FBM_Super} for superdiffusive FBM. %Notably, for $d = 2$ there is a shift from the theory when $A \to 0$ \red{(needs to be explained)}. These distributions are also plotted for the unsegmented tracks (Fig. \ref{fig:kite_analysis}f), although here the fit is not as good as in the previous cases.

\begin{figure*}[t]
\centering
\includegraphics[width=0.7\linewidth]{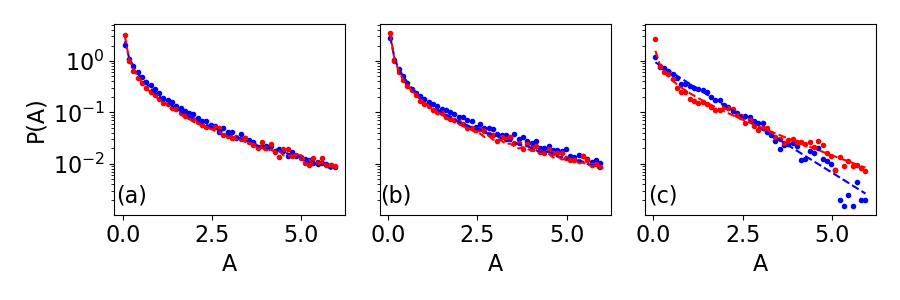}
 \caption{Distribution of single PSD around the average PSD for the three movement modes of a stork for $d = 1$ (red points) and $d = 2$ (blue points). In (a) the distributions are compared to simulations of subordinated FBM with $\alpha = 0.55$ and $\mathcal{H} = 0.40$ (red and blue dashed lines for $d = 1$ and $d = 2$ respectively) and similarly in (b) with $\alpha = 0.77$ and $\mathcal{H} = 0.70$. In (c) the distributions are compared to the theory for superdiffusive FBM given by Eq. \eqref{PA_FBM_Super} (red and blue dashed lines for $d = 1$ and $d = 2$ respectively). }
 \label{fig:storksPA}
\end{figure*}

\subsection{Stork}
The relatively low tracking frequency of the stork (one recorded point every five minutes) does not allow to identify and independently analyse potential searches, as these will include a very limited number of points (e.g., a search of 1 hour will include $\sim 12$ points). Thus, daily tracks are not segmented but rather considered as a single process, limiting the scope of the analysis.
Nevertheless, to properly account for the PSD for different behaviours, daily paths are clustered into subsets that represent distinct behaviours in a bird's life cycle~\cite{rotics2016challenges}. This is done using a Gaussian mixture model (GMM) \cite{scikitlearn} based on the logarithm of the maximum displacement ($\Delta X$) that the stork performs during each day (Fig. \ref{fig:storksGMM}a). We have found three significant movement modes which are highly correlated to the time of year: the mode with $\average{\Delta X} = 3$ km occurs primarily ($> 80\%$) between April and July, the mode with $\average{\Delta X} = 33$ km occurs primarily ($> 80\%$) between September and February, and the mode with $\average{\Delta X} = 213$ km occurs primarily ($> 87\%$) during well-known migratory periods~\cite{rotics2016challenges}. We thus refer to these three modes as breeding, wintering and migrating respectively and perform the analysis below for each subset of daily paths separately. Note that the GMM identified another mode with $\average{\Delta X} = 0.4$ km. However, as this mode includes a small number of daily paths (60 days out of $\sim 2500$), we discard it from our analysis.

For the breeding ensemble (966 days, Fig. \ref{fig:storksGMM}b) we found $\average{S} \sim t_m^{-0.45} f^{-1.9}$, which is consistent with subordinated FBM with $\alpha = 0.55 \pm 0.09$ and $\mathcal{H} = 0.40 \pm 0.05$, i.e., a non-stationary process with near $1/f^2$ noise.
For the wintering mode (1068 days, Fig. \ref{fig:storksGMM}c)  $\average{S} \sim t_m^{0.07} f^{-2.10}$, i.e., the noise is approximately $1/f^2$ noise and we find no significant ageing patterns. Importantly, the fact that we measure exponents $\beta = 2$ and $z = 0$ in Eq. \eqref{S_scaling_form} does not entail that the process is Brownian, in accordance with the results of Ref. \cite{vilk2021unravelling}. On the contrary, subordinated FBM with any $\alpha$ and $\mathcal{H}$ that fulfil $2\mathcal{H}\alpha = 1.07$ is a plausible model, as it is consistent with  $\average{S} \sim t_m^{0.07} f^{-2.10}$ and Eq. \eqref{PSD_FBM_CTRW}. Here, the knowledge of the analytical scaling of the PSD [Eq. \eqref{PSD_FBM_CTRW}] is not sufficient to determine the values of both $\alpha$ and $\mathcal{H}$ independently, and complementary analysis is crucial. For instance, based on the conclusions of Ref. \cite{vilk2021unravelling} one can determine $\mathcal{H} \simeq 0.7$ and $\alpha \simeq 0.77$, which is consistent with the above relation. Below, we show that the distribution of the single PSD further supports this.
% \red{As mentioned above, it is likely that a higher resolution tracks during both breeding and wintering would reveal behaviour specific details, that are obscured by analysing daily tracks.} %could be modelled by a subordinated FBM, and comparing to Eq. \eqref{PSD_FBM_CTRW} we find that $\alpha H = 0.56$ where $H > 0.56$.
For the migrating mode (480 days, Fig. \ref{fig:storksGMM}d) we get $\average{S} \sim t_m^{1.27} f^{-2}$, which is consistent with the ballistic motion of a migrating stork. Here, the stork migrates approximately in a straight line towards a stationary target (breeding and wintering grounds for spring and fall migrations, respectively), thus accounting for the rapid increase in amplitude (Fig. \ref{fig:storksGMM}d, inset). Notably, the fact that the scaling with the measurement time is steeper than linear implies that, in addition to the highly correlated nature of this process, the animal is also accelerating, which is again consistent with the complementary analysis in Ref. \cite{vilk2021unravelling}.\footnote{In particular, L{\'e}vy walks, that are often implicated as random search mechanisms for animals, were ruled out as governing process behind these data \cite{vilk2021unravelling}.} Indeed, storks roost in stopover sites during the night and tend to depart in the late morning, when soaring conditions improve, facilitating faster flights at lower energy costs~\cite{horvitz2014gliding}.

In Fig. \ref{fig:storksPA} we plot the distribution of the single PSD around its average for breeding, wintering and migration in one and two dimensions. For breeding we compare these distributions to simulations of subordinated FBM with $\alpha = 0.55$ and $\mathcal{H} = 0.40$ (see above). Similarly, for wintering we compare to subordinated FBM with $\alpha = 0.77$ and $\mathcal{H} = 0.70$. Here, the predicted distributions \eqref{PA_BM} for BM for $d=1,2$, can be shown to be far from the empirical distributions, entailing that this process is not Brownian. For both the breeding and wintering modes the empirical distributions show good agreement with simulated subordinated FBM. Note that although the ageing properties differ between these modes (see insets of Fig. \ref{fig:storksGMM}b-c), the distributions of the single PSD (Fig. \ref{fig:storksPA}a-b) are similar, suggesting some kind of unifying features of the underlying dynamics of the stork during breeding and wintering, see Ref. \cite{vilk2021unravelling} for further discussion. Finally, for migration (Fig. \ref{fig:storksPA}c) we find that the distribution is in excellent agreement with the theoretical distribution for superdiffusive FBM given by Eq. \eqref{PA_FBM_Super}.  %\red{ What can these tell us with lack of a theory? }

% \subsection{Vulture}

\subsection{Ageing effects}
\begin{figure}[t]
\centering
\includegraphics[width=1.\linewidth]{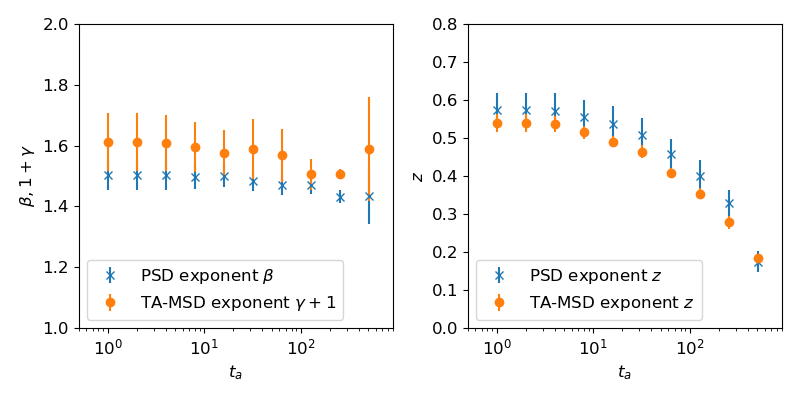}
 \caption{Exponents $\beta$ and $z$ in $\average{S} \sim t_m^{-z}f^{-\beta}$ [Eq. \eqref{S_scaling_form}, blue crosses] and exponents $\gamma$ and $z$ in $\average{\tasd} \sim t_m^{-z}\tau^{-\gamma}$ (orange circles) as measured for the kite's trajectories, as a function of the ageing time $t_a$. In (a) the error bars represents variations between different measurement times, while in (b) they represent variations between different frequencies $k$. }
 \label{fig:ageing}
\end{figure}

In many applications, the initial measurement time (the time from which the process is first observed) is different from the initiation time of the process, where the latter is often unknown. For nonergodic processes, such a discrepancy can make an anomalous process \textit{appear} closer to normal diffusion, as the observed statistics of the initial unmeasured period can be distinct from the statistics of the measured period  \cite{barkai2003aging, schulz2013aging, schulz2014aging, song2018neuronal,leibovich20171}. We denote the time between the initiation of the process and the beginning of the measurement by $t_a$, and refer to it as ageing time \cite{schulz2014aging, metzler2014anomalous}. Here, we measure the exponents $\beta$ and $z$ in Eq. \eqref{S_scaling_form} for the kite during ARS, as a function of $t_a$, see Fig. \ref{fig:ageing}. We vary $t_a$ between 1 and $512$ by discarding all points occurring at $t< t_a$ for each $t_a$ and obtain the PSD from the remaining points in the range $[t_a, T]$ (such that $t_m = T - t_a$), $T$ being the time since the \textit{unknown} initiation of the process. Notably, searches typically last $40$ min ($= 2400$ s), which is significantly longer than any choice of $t_a$.

For the PSD of the searching kite, we find that variations in $\beta$ are relatively small, with $< 10 \%$ difference between its value at $t_a = 0$ and its value at $t_a = 512$. This entails that ageing the trajectory (starting the measurement at increasingly long lag time $t_a$ instead of the initiation time of the process) does not affect the power-law shape and $1/f$ noise.  In contrast, the exponent $z$ significantly changes with $t_a$ ($> 70 \%$), meaning that ageing the trajectory changes the non-stationary properties of the system. For instance, if the process is modelled with subordinated FBM (Sec. \ref{sec:resultskite}), analysis of the PSD can give different values of $\alpha$, depending on $t_a$. This effect has important implications for ecologists, as it is often the case that the initiation time of a motion and the initial measurement time are not the same. Importantly, similar ageing effects can appear in the analysis of the EA-TA-MSD, as can be seen by comparing Eqs. \eqref{PSD_FBM_CTRW} and \eqref{MSD_FBM_CTRW}. In Fig. \ref{fig:ageing} we show a remarkable agreement between the two methods (PSD and MSD) with respect to the ageing analysis of the empirical tracks.

\section{Discussion}
In this work we analysed the PSD of ecological movement data. By segmenting and clustering the movement of a kite and a stork we are able to classify different behavioural modes based on the properties of the PSD.
For the kite we found different characteristics for bounded searches and commutes. The PSD of the former displays $1/f$ noise and strong ageing, both indicative of a non-stationary process and long-range correlations, as highlighted by comparison to known theoretical results and simulations of subordinated FBM. In contrast, for commutes, the process shows $1/f^2$ noise and an amplitude that increases with the measurement time, suggesting superdiffusive movement patterns of the commuting kite. These results are in agreement with the conclusions of Ref. \cite{vilk2021unravelling} for both the search and  commute modes, strengthening the validity of the models suggested above. Comparing these results to the analysis of unsegmented tracks highlights the improved insight one can obtain by properly identifying behavioural modes.

For all movement modes of the stork we find approximately $1/f^2$ scaling of the PSD; however, the ageing exponent $z$ significantly varies between modes. While breeding is a non-stationary process, wintering appears stationary as indicated by the exponent $z$, and migrating yields a PSD that increases with time. Although this analysis may be limited by the resolution of the data, as discussed above, we used known theoretical results and simulations to show that these daily movement patterns can be modelled as a subordinated FBM (breeding and wintering) or as an FBM (commuting).
The analysis performed here highlights the importance of considering the ageing exponent of the data and not solely the $1/f$ scaling. While the three modes of the stork only slightly differ in their $\beta$ values and may appear Brownian, they significantly differ in the ageing exponent. In particular, the breeding periods displays clear non-stationary patterns with decreasing diffusivity over time (see also~\cite{vilk2021unravelling}).

Although the time series for the stork are relatively short ($\sim 110$ points for each day), the analysis presented in Fig. \ref{fig:storksGMM} spans multiple decades in frequency. This is in contrast to the analysis performed in the temporal domain in Ref. \cite{vilk2021unravelling} where the power law scalings of the MSD were local and differed significantly between temporal scales. This indicates that the PSD is typically more robust, allowing for more accurate estimates of the underlying process. However, we stress that the PSD analysis used here [specifically Eq. \eqref{PSD_FBM_CTRW}] does not generally give a unique set of model parameters. As shown above for the wintering stork, the PSD exponents can be consistent with a range of interconnected values of $\alpha$ and $\mathcal{H}$. Moreover, one could mistakenly assume that the process is close to Brownian. Here, complementary analysis is vital for correct inference. For instance, analyses based on the amplitude scatter of the time-averaged MSD \cite{metzler2014anomalous,he2008random}, the p-variation method \cite{weron,weron1}, or the first-passage time distribution \cite{fpt,fpt1} are useful statistical observables for single particle tracking data.
Nevertheless, our analysis demonstrates that the PSD method provides vital information in the analysis of movement data.

% The method discussed here is complementary to other methods, such as... The PSD is typically more robust to noise, allowing for more exact estimates of the underlying process. (look at Eli's paper for this discussion).

%What theory is missing? Perhaps the distribution of the PSD for CTRW. Also a theory for ageing. Data driven analysis.
Finally, our results highlight possible future directions, particularly in the study of the subordinated FBM and the CTRW formalism. First, to better quantify the empirical distributions found in Figs. \ref{fig:kite_analysis_PA} and \ref{fig:storksPA}, a theoretical study of distribution of the PSD for CTRW and subordinated FBM is required.
Second, it has recently been shown that anomalous transport processes which do not obey the predictions of the Gaussian central limit theorem, can be decomposed into three individual effects (exponents); non-stationarity, temporal correlations and extreme events~\cite{chen2017anomalous, aghion2021moses, vilk2021unravelling,thapanjp}. It can thus be interesting to test whether the PSD method can be extended as an independent method to extract these exponents (the PSD analysis we use here only provides two independent exponents). To this end, it would also be useful to compare the insights extracted directly from the ageing PSD of the data, with that obtained using more data-driven approaches, such as machine-learning~\cite{zhang2020deepmmse,ANDI2021,lurig2021computer, munoz2020single, granik2019single, kowalek2019classification}, or Bayesian maximum likelihood analysis~\cite{thapa2018bayesian}. Finally, following the empirical evidence that an ageing time can drastically affect the measured exponents, a general ageing theory for both CTRW and subordinated FBM is needed.

\section{Acknowledgements} 
For fieldwork and technical assistance we thank Y. Orchan, R. Shaish, A. Levi, S. Rotics and M. Kaatz. RN acknowledges support from JNF/KKL grant 60-01-221-18 and DIP (DFG) grant NA 846/1. RN also acknowledges support from The Minerva Foundation, the Minerva Center for Movement Ecology and the Adelina and Massimo Della Pergola Chair of Life Sciences.  
RM acknowledges the German Science Foundation (DFG) for support within grant ME 1535/12-1.
OV and MA acknowledge support from the ISF grant 531/20. 
MA also acknowledges Alexander von Humboldt Foundation for an experienced researcher fellowship.

\bibliography{refs}

\end{document}